# Boosting the transparency of metallic SrNbO$_3$ through Ti doping

Shammi Kumar[1,a)], Liang Si[2,3,4, b)], Karsten Held[4, c)], Sankar Dhar[1], Rakesh Kumar[1] and Priya Johari[1,a)]

[1] *Department of Physics, School of Natural Sciences, Shiv Nadar Institution of Eminence (Deemed to be University), Gautam Buddha Nagar, Uttar Pradesh 201314, India*
[2] *School of Physics, Northwest University, Xi'an 710127, China*
[3] *Shaanxi Key Laboratory for Theoretical Physics Frontiers, Xi'an 710127, China*
[4] *Institute of Solid-State Physics, Vienna University of Technology, 1040 Vienna, Austria*
a) **Authors to whom correspondence should be addressed**: priya.johari@snu.edu.in, sk657@snu.edu.in
b) siliang@nwu.edu.cn
c) held@ifp.tuwien.ac.at

**Abstract**

In recent years, various materials have been developed to reduce the reliance of industries on Indium, a primary component of transparent conducting oxides (TCOs) used in the current generation of devices. The leading candidates for indium-free TCOs are strontium vanadates, niobates, and molybdates—strongly correlated perovskite systems that exhibit high intrinsic electrical conductivity and optimal transparency. In this work, we focus on the strontium niobate (SrNbO$_3$) thin films and manipulate its optical conductivity by Ti doping which shifts the plasma frequency and reduces electronic correlations. This allows us to achieve a low resistance of SrNb$_{1-x}$Ti$_x$O$_3$ (x = 0 - 0.5) thin films while maintaining a high transparency in the visible light region. We obtain the optimal figure-of-merit (FOM) of 10.3 (10$^{-3}$ Ω$^{-1}$) for x = 0.3. This FOM significantly outperforms the optoelectronic capabilities of Tin-doped Indium oxide (ITO) and several other proposed transparent conductor materials. Our research paves a way for designing the next generation of transparent conductors, guided by insights from density-functional theory (DFT) and dynamical mean-field theory (DMFT).

**Introduction**

The demand for transparent electronics has surged in recent decades, driven by the widespread use of consumer electronics like smartphones, televisions, LEDs etc.[1–3]. It is also one of the primary requirements for the fabrication of solar cells[4-7]. At the heart of all such display devices are thin film transparent conducting electrodes which are, as the name suggests, simultaneously transparent to the visible light as well as electrically conducting. The ideal transparent conductors are ones which have a high band gap ($E_g$ > 3.1eV), exhibit high transparency (T > 85%) in the visible light range and, at the same time, a high conductivity ($\sigma$ >10$^3$ S/cm)[8,9]. Among several candidates, Tin-doped Indium oxide (ITO) has become the standard transparent conductive oxide (TCO) globally due to its outstanding optoelectronic performance. It has a high transparency (>85%) and a low sheet resistance (10-100 Ω □$^{-1}$), making it suitable for a wide variety of optoelectronic devices[1,5,10]. However, due to the high cost and scarcity of Indium, a better alternative to reduce our reliance on ITO is urgently needed[11]. Significant progress has been made in this area, and several alternatives are currently being explored. One of the primary candidates to replace Indium based TCOs is correlated perovskite oxides, such as SrVO$_3$ (SVO), SrMoO$_3$ (SMO), and SrNbO$_3$ (SNO), which are correlated metals that exhibit low resistivity (30-200 μΩ cm$^{-1}$)[12]. Their transparencies are quite low (40-70%) and can be explained by the plasma frequency $\omega_p = \sqrt{\frac{(4\pi e^2 n)}{(\epsilon_r m^*)}}$, where $e$ is the electronic charge, $n$ the density of conduction electrons, $m^*$ the effective mass, and $\epsilon_r$ the static screening[13]. All these oxides have a relatively high $n$ (~10$^{22}$ cm$^{-3}$), correspondingly the plasma frequency is expected to be quite high. Moreover, for correlated systems, the effectively mass renormalization factor $Z^{-1} = m^*/m_b$, where $m_b$ is the band mass obtained from DFT calculations, must be taken into account to better understand the observed trends of the plasmon frequency[14]. Here Z = 1 for non-interacting electrons and Z = 0 for fully localized electrons, i.e., Mott-insulator. Light that has a frequency smaller than $\omega_p$ is mostly reflected back whereas light with a frequency greater than $\omega_p$ is able to penetrate deeper into the sample, leading to the optical transparency. The visible light has the energy range of 1.75 - 3.10 eV, meaning the plasmon frequency of a given material should be smaller than 1.75 eV to be transparent to the visible spectrum. Among the above-mentioned perovskites, only SVO has a $\omega_p$ (~1.3 eV) which lies below the visible range making it an effective transparent conductor[11,15,16]. However, the small $\omega_p$ of SVO occurs due to a large band mass-enhancement (Z ~ 0.33, $m^* = 3m_b$) that, reduces the electronic lifetime $\tau$ (enhances the scattering rate $\tau^{-1}$); as a result of which it has still a quite high absorbance above 2.25 eV[17]. This scattering is further increased at ~2.7 eV, where interband transitions dominate which results in an enhancement in the absorption, thereby reducing the transparency[13,17]. The potential for usage of SVO as a transparent conductor is limited due to this shortcoming.

On the other hand, the $\omega_p$ of SMO (Z ~ 0.48) lies within the visible range (~ 2 eV) which decreases the transparency at low energies in the visible spectrum, limiting their use till date[14,18–20]. Several notable strategies have been proposed recently to shift either $E_g$ or $\omega_p$ of SVO and SMO to enhance their transparency, such as, thickness modulation, A and B site cationic substitution, or strain modulation[21–26]. In particular, the cationic substitution of V by Mo, performed by Mohammadi *et al*. has been successfully reported to regulate $\omega_p$ as well as $E_g$ and established $SrMo_{1-x}V_xO_3$ as an alternative to the current TCOs[27].

In this work, we focus on another material, SNO, and investigate its potential as a TCO. SNO has a larger transparency range than SVO and SMO, and has fantastic properties for plasmonics, photocatalysis, two dimensional electron gases (2DEGs), and metal-insulator transition (MIT) device[28–38]. SNO is a large (optical) band gap material ($E_g$ ~ 4.1- 4.5 eV) with the reported resistivity of 50-200 μΩ-cm making it a promising candidate to replace ITO[29,39,40]. However, the transparency of the metallic SNO is quite low (40-70%), due to the relatively high $\omega_p$ (1.8 - 2.0 eV) as well as the free carrier absorption (FCA), which impedes its intended application as a TCO[29,39,40]. A significant amount of work has been dedicated to enhancing the optical transmittance of SNO through various techniques. For example, Roth et al. grew SNO thin films of different thicknesses (t ~ 10 - 55 nm) on $(La_{0.3}Sr_{0.7})(Al_{0.65}Ta_{0.35})O_3$ (LSAT) and successfully improved the transparency to 86% by using sputtering deposition technique[41]. However, the cost of this is the sheet resistance of thin films increased from 74 Ω to 370 Ω, which is unfavorable for TCO. Another technique to increase transparency is by inducing off-stoichiometry[8,39,42,43]. However, this method has also only succeeded in raising optical transparency to 75%, with a slight trade-off in conductivity. Recently, Jeong et al. were able to optimize the growth conditions of SNO on $DyScO_3$ (110) substrates, to achieve a reported $R_s$ of 10 Ω and maximum transparency, $T_{max}$ ~87%, marking a significant increase in the figure of merit[44]. However, the transparency falls rapidly at lower energy ranges (~2eV), possibly due to $\omega_p$. These various approaches highlight a significant inconsistency in finding an effective method to optimize the optoelectronic properties of SNO. Additionally, other theoretical strategies, such as strain tuning and cationic substitution, have been reported to have minimal impact on the plasmon frequency and, consequently, on the transparency of SNO[22,24].

A promising approach to enhancing the properties of SNO was recently proposed by Si et al.: Ti ($d^0$) doping of SNO (STNO) was identified as a viable strategy to improve the optoelectronic properties of SNO[17]. Ti doping has a dual effect: First, it reduces the number of free carriers in STNO, reducing $\omega_p$ and FCA. Secondly, going away from an integer Nb $d^1$ configuration, correlation effects are reduced. This results in a larger, less strongly renormalized quasiparticle weight (Z), which would typically increase $\omega_p$. However, the overall trend is a decrease $\omega_p$ due to Ti (hole) doping. Reduced correlations also lead to less scattering and longer carrier lifetimes, which are crucial for enhancing conductivity. Together, these effects may shift the absorption edge below the visible light spectrum, while maintaining high conductivity in the material.

Here, we verify the direct and simple approach proposed by Si *et al.* on the SN$_{1-x}$Ti$_x$O system, by varying the Ti concentration from 0 - 0.5. In our DFT+DMFT calculations[45], we observe that 25-50% Ti doping does not significantly affect the electronic conductivity but successfully shifts $\omega_p$ below the visible spectrum. We then confirm this experimentally by growing epitaxial thin films of SN$_{1-x}$Ti$_x$O ($t \sim 70$ nm) on LaAlO$_3$ (LAO) substrates (schematic Figure 1(a)). It is seen from the photographs of thin films in Figure 1(b) that, as the composition of Ti is increased, the films become increasingly transparent, even almost completely transparent at higher Ti composition (x = 0.5). In Figure 1(c), we plot the FOM of these films and find that it increases drastically upon increasing the concentration of Ti in these films up to x = 0.3 and decreases slightly for x = 0.5. As we mentioned earlier, the optoelectronic properties are highly tunable by growth conditions as shown by the recent improvement in FOM of SNO[44]. The same applies to SN$_{1-x}$Ti$_x$O thin films, which are highly sensitive to growth conditions. The FOM can be further enhanced by optimizing factors such as substrate selection, film thickness, $p$O$_2$ levels, and growth temperature. To show this variation in our results, we also display the highest achieved FOM (shown by star) for individual concentration by varying the growth oxygen partial pressure ($p$O$_2$) and the thickness in Figure 1(c). To simplify the optimization process, we focus on improving transparency solely through Ti substitution, while maintaining consistent growth conditions for all thin films in this study.

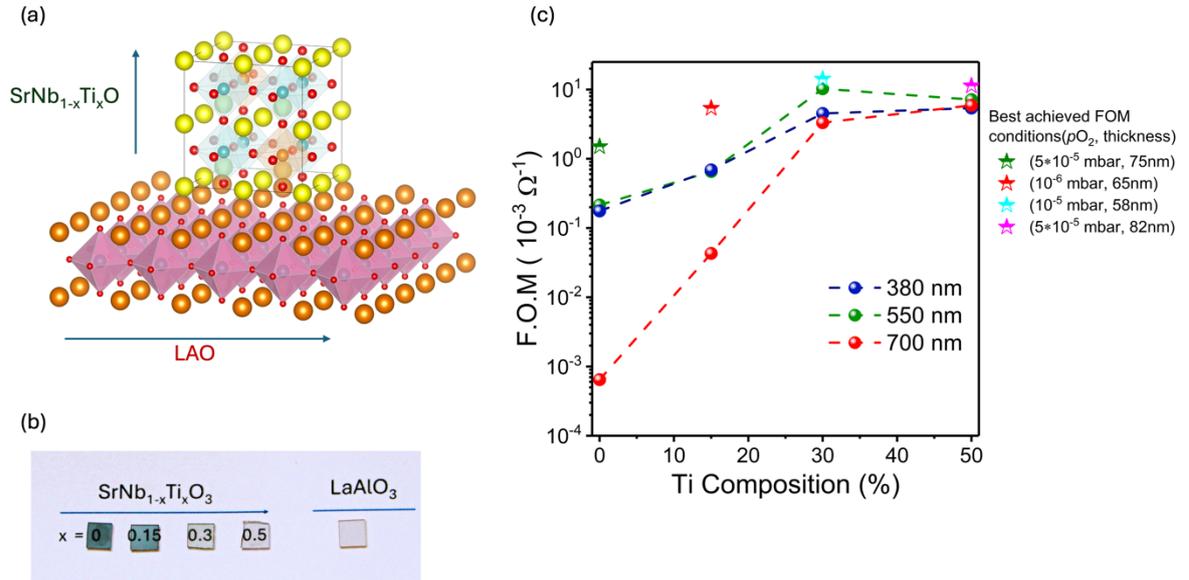

Figure 1: (a) Schematic of SN$_{1-x}$Ti$_x$O films grown on LAO. (b) Photograph of the thin films as a function of doping showing an enhancement in the transparency upon Ti doping; LAO is given on the right as a reference. (c) FOM of thin films of ~70 nm at three wavelengths covering the entire visible range. The FOM is highly improved, especially for 30-50% Ti doped SNO (solid circles). The half-filled stars represent the best achieved FOM for each individual concentration obtained for the conditions given on the right. Unless explicitly mentioned, the thin films in Figure 1(b), (c) (solid circles) and throughout the manuscript are grown under the same conditions $p$O$_2$=10$^{-5}$ mbar and $t$ ~70nm.

# Results

## Theoretical Calculations

### *(a) Electronic Properties*

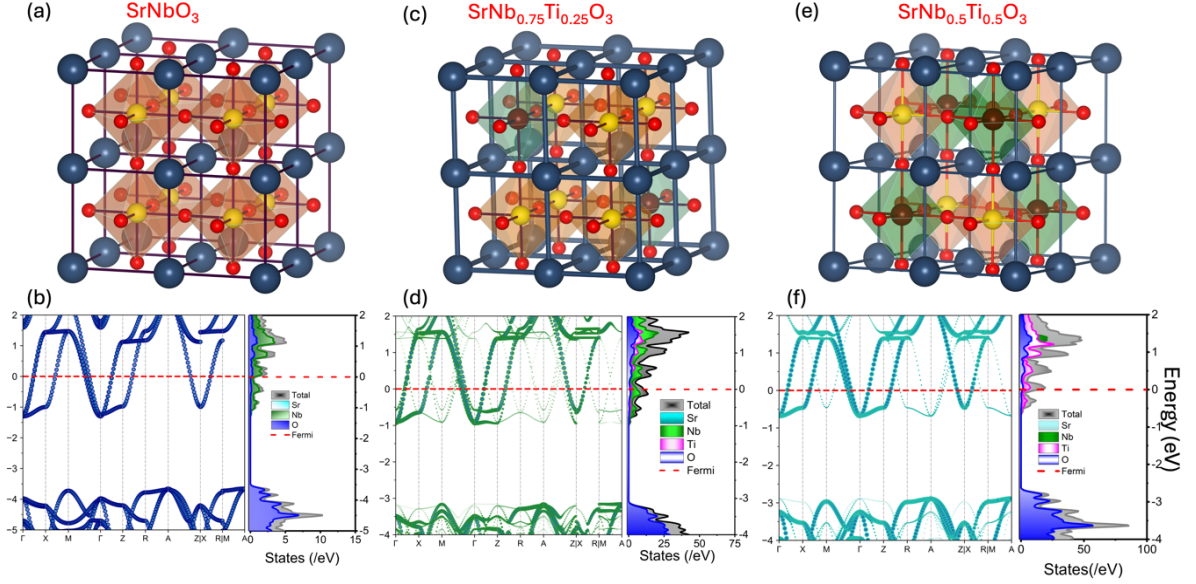

Figure 2: Band structure and partial density of states (bottom) for different supercells (top), and unfolded band structures of the $SN_{1-x}T_xO$ system for (a,b) x = 0, (c,d) x = 0.25, and (e,f) x = 0.5. The yellow polyhedral represents the $NbO_6$ octahedra and $TiO_6$ octahedra are shown by the green polyhedral. The Fermi level shifts downwards indicating a decrease in the band gap but still remains well within the conduction band indicating a metallic-like conductivity.

To investigate the impact of Ti doping on the electronic features of the $SN_{1-x}T_xO$ system, we perform DFT calculations. For these, we first construct a 2×2×2 supercell of SNO and then substitute 0, 2, and 4 Nb atoms with Ti atoms, resulting in effective compositions of x = 0, 0.25, and 0.50 in the $SN_{1-x}T_xO$ system. Band structure calculations are then performed to obtain the crystal structures with lowest total energy. This setup aims to reduce the strong scattering resulting from the disorder caused by the introduction of doped atoms. In order to study the band changes induced by the doping of Ti atoms, we employ the band unfolding methodology as outlined by Popescu and Zunger to accurately determine the electronic properties[46–48].

Figure 2 (a), (c), and (e) represent the supercell structure of $SN_{1-x}T_xO$ used for the calculation of the band structure, while Figure 2 (b), (d), and (f) show the full and unfolded band structure and partial density of states (*p*DOS) for x = 0, 0.25, and 0.5, respectively. Here, the $NbO_6$ octahedra are shown in yellow color and the $TiO_6$ octahedra in green; the Fermi level is set to 0 in each case. From the *p*DOS, we see that the valence band maxima (VBM) in each case is dominated by O *2p* orbitals, while the conduction band minima (CBM) is mainly composed of Nb *3d* orbitals for x = 0 and 0.25, whereas for x = 0.5 both Nb and Ti *3d* orbitals have equal contribution to the conduction band. Upon doping Ti into the system, the Fermi level shifts downward within the conduction band, indicating a reduction in the number of free carriers (electrons) in the system. Also, the position of CBM changes from ~ -1.3 eV to ~ -0.8 eV from x = 0 to 0.5, indicating a Fermi level shift of ~0.5 eV. Despite this huge shift, the Fermi level remains well within the conduction band, thus indicating a metallic character even for a large amount of Ti doping. From the unfolded band structure, one can see clearly that the nature of the band gap remains the same, i.e., the interband transition (O *2p* → Nb/Ti *3d*) is indirect for all cases (along the A→Γ direction).

As mentioned above, due to the scattering potential as well as the lost translational symmetry, the band structure unfolding results in several spurious bands for the doped compounds, owing to the incommensurability between the supercell and primitive cell and can be ignored. With these caveats, it is seen that the bands around the Fermi level retain their parabolic shape in all cases, which indicates that the effective mass and hence the mobility of the carriers is not expected to vary much from the mobility of SNO. Moreover, due to the shift in the Fermi level, the band gap ($E_g$) between the fully filled levels in the valence band to the conduction band decreases with increasing Ti doping. This should not come as a surprise, since the end products of $SN_{1-x}T_xO$ (x = 0 and 1) have $E_g$ = ~ 4.1 eV (SNO) and 3.2 eV (STO), respectively. It can thus be expected that for intermediate composition, $E_g$ remains between these values. This is verified by the band structure properties shown in Figure 1 (see supplementary information Figure S1 for the band structure of STO).

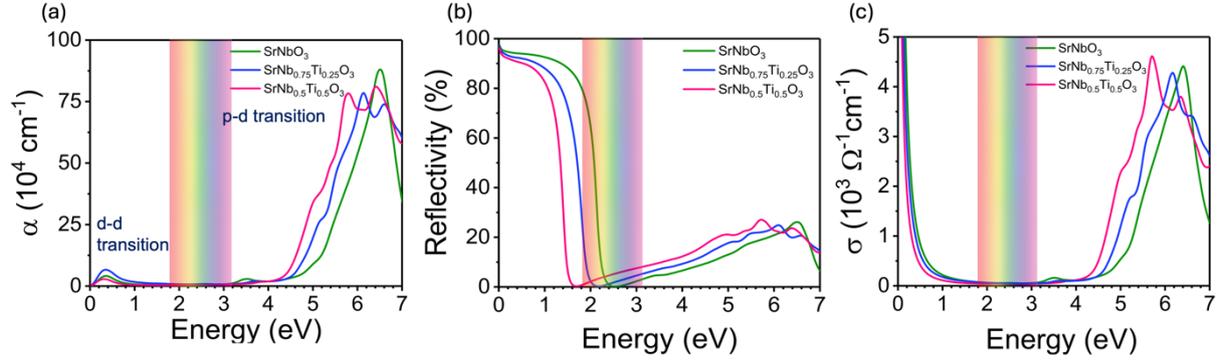

Figure 3: (a) Absorption coefficient, (b) reflectivity, and (c) optical conductivity for $SN_{1-x}T_xO$ in DFT (x = 0, 0.25, and 0.5). Ti doping shifts the absorbance and reflectivity minima below the visible range thus predicting an enhancement of transparency in the visible range. In (a), *p-d* transition represents the interband transition and *d-d* transition shows intraband transitions.

### *(b)   Optical Properties*

While SNO has a large band gap (~ 4.1 to 4.5 eV), the transparency of SNO is not optimal due to strong intraband absorption and high $\omega_p$. The intraband transition in $SN_{1-x}T_xO$ cannot be explained by band structure studies alone, as it only provides meaningful insight for edge-to-edge transition and does not consider other effects such as Hubbard bands at higher energies, multiband effects etc. Hence, in order to get a deeper insight into the optical properties, we further calculate the frequency-dependent optical properties by DFT to simulate the behavior of light across the UV-Vis-NIR spectrum to take care of these effects.

Figure 3 (a) presents the DFT-calculated absorption coefficient of the $SN_{1-x}T_xO$ system. While the absorption coefficient (α) for the entire $SN_{1-x}T_xO$ system is suppressed in the visible region (~$10^5$ cm$^{-1}$), it increases quite significantly on both sides of the visible spectrum. At high energies (≳ 4.0 eV), α increases quite significantly which is due to the the *p-d* transition as shown. This is the familiar edge-to-edge transition (or $E_g$) which we get from band structure calculations (Figure 2) and which are very far from the visible region. However, an enhancement of α also occurs on the lower energy side which is marked as *d-d* transition in the figure and is the intraband absorption. As can be seen from Figure 3(a), doping with Ti mitigates this problem and enhances the transparency of the material. The effect of large doping is seen in the visible region as well, since the α of $SN_{0.5}T_{0.5}O$ is very low compared to the other two cases. Another important understanding of the optical property of $SN_{1-x}T_xO$ is seen from the reflectivity data in Figure 3(b): the reflectivity drops to zero at ~ 2.4 eV for SNO and increases suddenly at lower energies. When doping with Ti, the reflectivity edge begins however to gradually shift towards the lower energies until finally at x = 0.5 this reflectivity edge shifts entirely out of the visible spectrum, making $SN_{0.5}T_{0.5}O$ highly transparent as compared to SNO. The shift of the absorption edge to lower energies as seen in the optical conductivity plot Figure 3(c) is consistent with the DFT band structure of Figure 2, where, as the concentration of Ti is increased, the CBM shifts upwards resulting in a smaller *p-d* transition energy.

The DFT obtained optical properties are helpful to gain insight into the effect of doping on the optical properties of $SN_{1-x}T_xO$. However, due to strong correlations in the *d* orbitals of SNO, the simplistic Drude model used to calculate the optical properties does not fully agree with the experimental results. Further, including correlations by DMFT, in particular the mass enhancement ($Z^{-1} = m^*/m_b$) and scattering rate ($\tau^{-1}$) is essential[17].

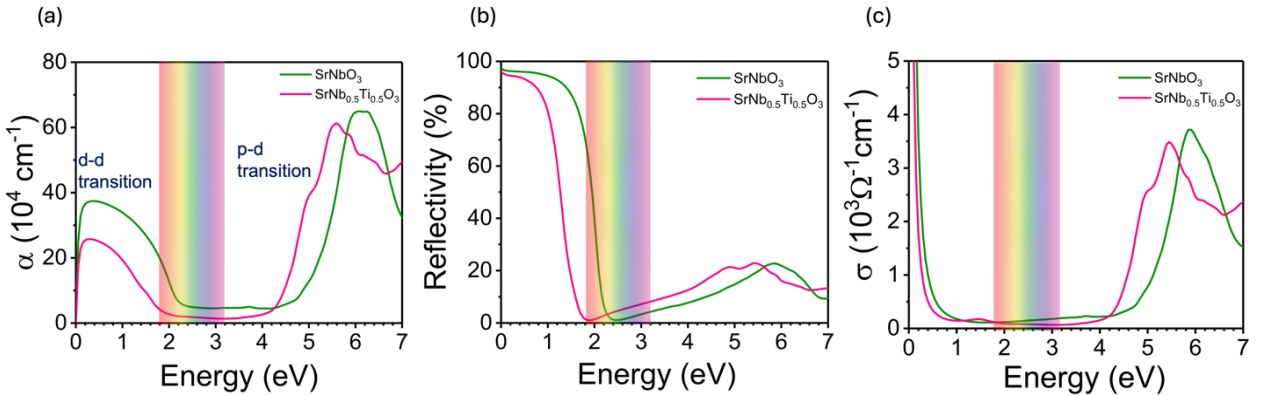

Figure 4: DMFT obtained optical properties of $SN_{1-x}T_xO$: (a) absorption coefficient and (b) reflectivity. In both (a) and (b), the absorbance and reflectivity minima have a finite value, as opposed to the DFT obtained results, but are shifted below the visible spectrum, showing the same trend as DFT. (c) DMFT calculated optical conductivity.

In Figure 4 we present the DMFT obtained optical properties for x = 0 and 0.5 to get an accurate description of the inter- and intraband transitions across the spectrum. Comparing Figure 3(a) and 4(a), it is seen clearly that the inclusion of electron correlation results in enhanced scattering at low energies which further increases the absorption coefficient at lower energies. When comparing the DFT and DMFT results with the experimentally obtained α (See SI-Figure S2), it is seen clearly that DMFT produces the correct behavior of α. Furthermore, the reflectivity minimum shown in Figure 4(b) is no longer zero, but finite, due to correlation-enhanced scattering at low energies The resulting optical conductivity in Figure 4(c) shows a non-zero conductivity in the visible region even at higher doping concentrations. Nonetheless, the general trend is similar to DFT, i.e., Ti doping shifts the absorption edge and reflectivity edge towards the lower energy side of the visible spectrum.

What does this mean in terms of TCO properties? For a start, the shift in $\omega_p$ results in the shift of the reflectivity and absorption edge below the visible spectrum. For a given material to be transparent to the visible light, its plasmon frequency should be lower than 1.75 eV and its band gap higher than 3.1eV. By performing DFT+DMFT calculations, we have shown that we can shift the plasmon frequency below the visible spectrum whilst keeping the band gap higher than 3.1eV by Ti doping of SNO. $SN_{1-x}T_xO$ is expected to exhibit a metallic-like conductivity even for larger Ti doping. Let us now turn to the experimental results to verify the theoretical results.

# Experimental results

## *(a) Structural properties:*

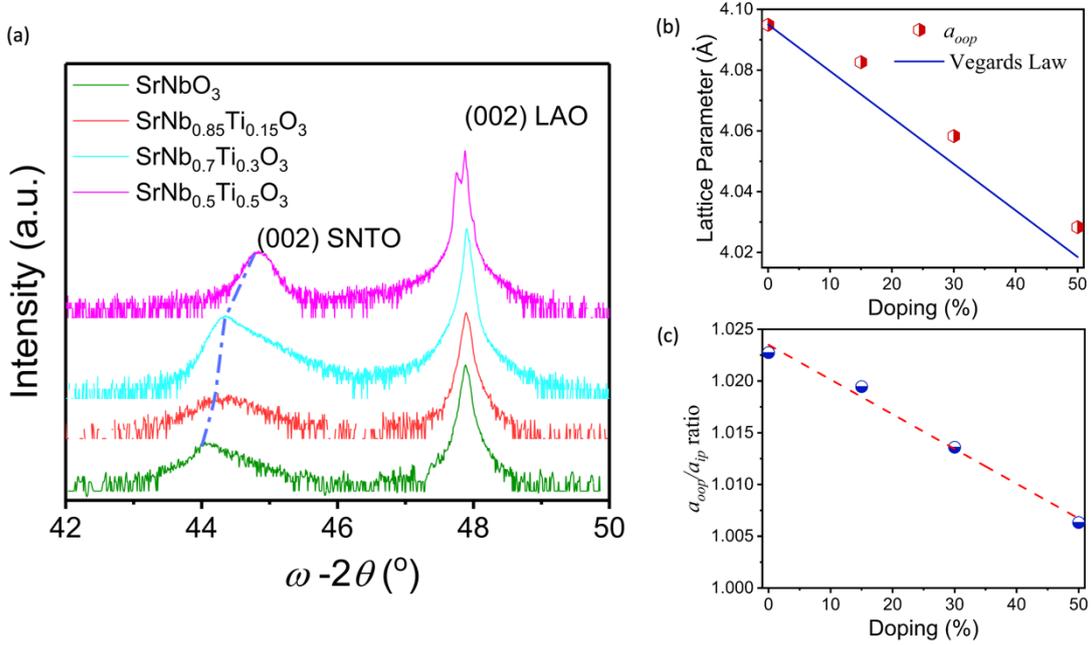

Figure 5: (a) HR-XRD spectrum of the (002) peak of $SN_{1-x}T_xO$ system for x = (0, 0.15, 0.3, 0.5) grown epitaxially on (001) LAO. The peaks corresponding to the film and substrate are indicated by a dashed blue line. A shift in peak towards higher $2\theta$ values is seen as the composition of Ti is increased in the system. (b) Change of the out-of-plane lattice parameter of the thin films as a function of Ti composition shows a shrinkage in the lattice parameters of the $SN_{1-x}T_xO$ system. Due to the compressive strain induced by LAO, $a_{oop}$ is always larger than the lattice parameters obtained from Vegard's law. (c) $a_{oop}/a_{ip}$ ratio, demonstrating that Ti doping reduces the compressive strain within the unit cell.

The structural characterization of the epitaxial $SN_{1-x}T_xO$ system (x = 0, 0.15, 0.3, 0.5) grown on (001) LAO is studied with the help of high-resolution x-ray diffraction (HR-XRD). In each instance, only the (001) family of peaks corresponding to the $SN_{1-x}T_xO$ system is observed, confirming the epitaxial nature of the thin films. In figure 5(a) we show the *ω-2θ* scans around the (002) diffraction peak for the thin films and the substrate to show the variation of peaks with Ti composition. The thin films are grown on (001) LAO single crystal substrates, with a lattice constant of 3.79 Å. As a result of the large lattice mismatch between the film and the substrate, a compressive strain is seen in all the films, which elongates the out of plane lattice parameter ($a_{oop}$). Further, when Nb (0.64 Å) is replaced by Ti (0.605 Å), the lattice shrinks due to the smaller size of the Ti ion. This corresponds to the peak shift of (002) $SN_{1-x}T_xO$ to a higher 2θ value which increases as the amount of Ti increases in the composition. In Figure 5(b), we have calculated $a_{oop}$ as a function of composition of Ti on $SN_{1-x}T_xO$. The in-plane lattice parameter ($a_{ip}$) for all the films remains constant at 4.003 Å and, as seen in supplementary figure S3, the films are four-fold symmetric which implies that these films have a tetragonal structure. Moreover, we observe that there is a slight deviation from Vegard's law in all the cases ($a_{SNO}$ = 4.1Å, $a_{STO}$ = 3.94Å)[49]. It is known that both Nb and Ti ions have two valence states. When $Nb^{4+}$ is replaced by $Ti^{4+}$, the substitution is isovalent, whereas when two $Ti^{4+}$ ions are replaced by $Ti^{3+}$ and $Nb^{5+}$ ions it is heterovalent[50]. The valency of Nb and Ti is highly dictated by the growth condition and in the case of isovalent substitution the $a_{oop}$ should follow Vegard's law. Since our films do not follow Vegard's law ideally, it evidences that there is some intermixing of the different valence states amongst the Ti and Nb ions. These conclusions are consistent with the DFT and DMFT calculations, and we have shown in our previous work how the deposition condition affects the valency of Nb[43]. The $a_{oop}/a_{ip}$ ratio in Figure 5(c) shows that the tetragonal distortion of thin films is reduced as the content of Ti is increased. This is further complemented by Figure S3(d), in which we see that the full width half maxima (FWHM) of the rocking curve in the case of $SN_{0.5}T_{0.5}O$ is reduced as compared to that of pure SNO, indicating a better epitaxial matching with the substrate.

## (b) Electronic Properties

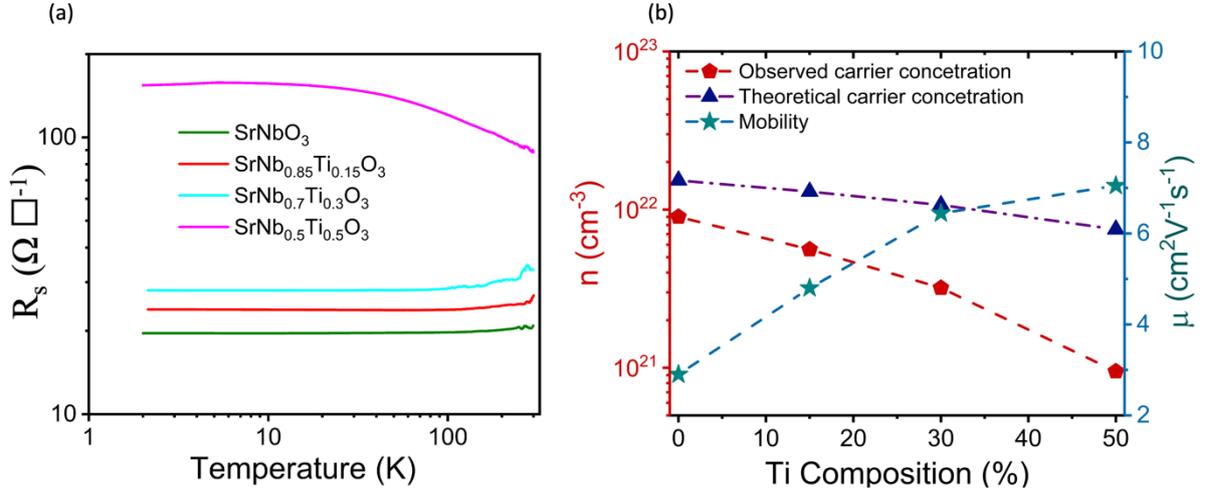

Figure 6: (a) Measured sheet resistance for different compositions of Ti in $SN_{1-x}T_xO$ thin films. The sheet resistance increases gradually as the amount of Ti increases. The films exhibit a metallic nature for x = 0 - 0.3 but show a semiconductor type of behaviour for x = 0.5. However, the sheet resistance remains below 100 $\Omega$ at RT (b) Experimentally obtained Hall mobility and carrier concentration for $SN_{1-x}T_xO$ thin films compared with theoretical carrier concentration.

The sheet resistance of the $SN_{1-x}T_xO$ thin films is shown in Figure 6 (a), demonstrating that all films are metallic. The room-temperature sheet resistance ($R_s$) increases from 18 to 89 $\Omega\, \square^{-1}$ as the concentration of Ti is increased; and the $SN_{1-x}T_xO$ films exhibit a metallic-type ($dR_s/dT > 0$) character up to x = 0.3 Ti doping but a semiconductor-like ($dR_s/dT < 0$) behaviour at x = 0.5. Since all the films are ~70-80 nm thick, this yields a resistivity value ranging from ~ 200-900 $\mu\Omega$-cm at room temperature (RT). In Figure 6(b), we show the calculated Hall mobility and carrier concentration for $SN_{1-x}T_xO$ thin films. For completeness, we have also included the theoretically calculated carrier concentration, using the formula $n^{theory} = n_{f.u}/V_{f.u}$. Here, $n_{f.u}$ represents the number of electrons per formula unit, $V_{f.u}$ is the volume and $n^{theory}$ is the theoretically calculated carrier concentration. It is seen that the observed carrier concentration is consistently less than the expected carrier concentration. We attribute this change to heterovalent mixing of $Nb^{5+}/Nb^{4+}$ and $Ti^{4+}/Ti^{3+}$ ions for our deposition condition. However, the mobility of the $SN_{1-x}T_xO$ shows a reverse trend to the carrier concentration. We observe that the mobility (μ) of the system increases unexpectedly from ~ 3 to 7 $cm^2V^{-1}s^{-1}$ which is closer to the range of mobilities reported earlier for $SN_{1-x}T_xO$ thin films[50]. This change in mobility cannot be explained easily as there might be several factors at play. One prospective explanation is the non-integer occupations of Ti $d^{0.58}$ and Nb $d^{0.42}$. This means weaker correlations between electrons, a reduced electron-electron scattering, and thus a higher mobility. A second factor increasing the mobility can be possibly attributed to the better epitaxy and reduced residual strain of the unit cell in Ti doped SNO system. This might suppress the scattering and increase the relaxation time of the electrons. In summary, all the films exhibit metallic character with a maximum sheet resistance of 89 Ω at RT for x = 0.5 which is consistent with the electronic properties obtained theoretically. We next perform the optical characterization to evaluate the performance of $SN_{1-x}T_xO$ as a TCO.

*(c) Optical Properties*

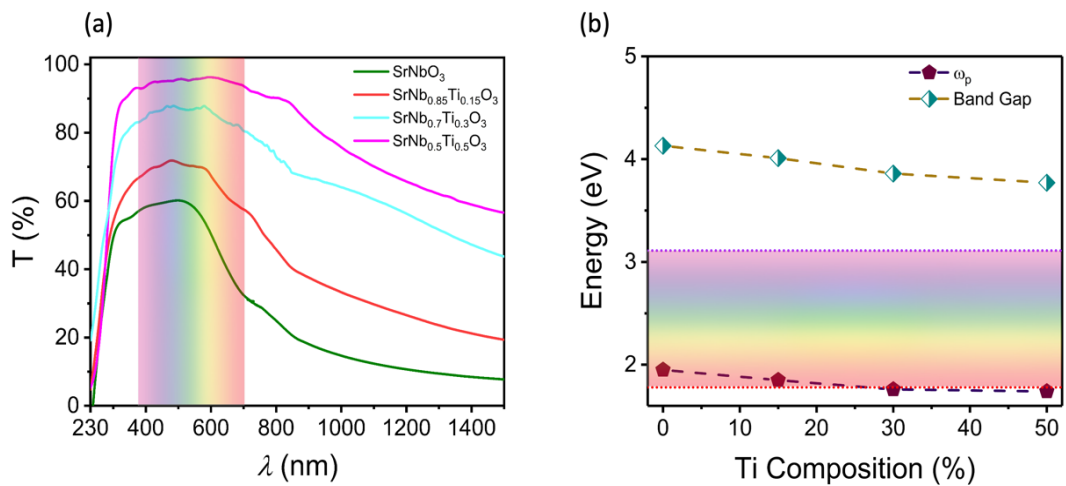

Figure 7: (a) Optical transmittance spectra of $SN_{1-x}T_xO$ thin films. The transmittance of the thin films increases from ~60% to 95% as x changes from 0 to 0.5. (b) Experimentally observed band gap and plasmon frequency as a function of Ti composition. The band gap remains well above 3.2 eV while decreasing the plasmon frequency below 1.75 eV.

Figure 7(a) shows the optical transmittance of $SN_{1-x}T_xO$ thin films. As we predicted theoretically, the optical transmittance of the thin films increases as the amount of the Ti is increased. For x = 0.3 and 0.5, the optical transmittance of the thin films reaches 87% and 95% at 550 nm (or 2.25 eV), the wavelength at which our eyes are most sensitive to. Moreover, it is seen that the transmittance of $SN_{1-x}T_xO$ in the UV range (~300 nm) is 80-90%, whereas in the IR range (~1000 nm) it is around 60-70% (for x = 0.3 and 0.5). These transmittances significantly improve upon previously reported results for such strongly correlated systems and confirm the prediction made by DFT+DMFT that we showed earlier.

In figure 7(b) we have calculated the band gap and the plasmon frequency of these thin films. The band gap of the thin films is calculated with the help of Tauc's plot by considering an indirect transition (as seen in band unfolding results). The Tauc's plot and the calculated band gaps are shown in Figure S4. The plasmon frequency of the thin films is extracted through variable angle ellipsometry (VASE) by considering the zero crossover of the real part of the dielectric function. The VASE data along with the real and imaginary part of the dielectric function is given in Figure S5. For x = 0.3, $\omega_p$ falls just inside the visible region, while for x = 0.5 it is completely outside. As a result of this, the optical transmittance of $SN_{0.5}T_{0.5}O$ is further improved compared to $SN_{0.7}T_{0.3}O$. The other condition for maintaining high transparency i.e., a band gap greater than 3.1eV, is observed across the entire doping range.

### (d) Figure of Merit

We finally evaluate the performance of $SN_{1-x}T_xO$ as transparent conductor by invoking Haacke's FOM formula[17]:

$$FOM = \frac{T^{10}}{R_s}.$$

Here T is the transparency and $R_s$ is the sheet resistance. Taking the values of $R_s$ and T at three different wavelengths, we cover the performance of $SN_{1-x}T_xO$ in the entire visible region. Figure 1(c) shows the FOM obtained for thin films at 380, 550 and 700 nm. It is seen that doping improves the optoelectronic properties of the thin films, especially at high wavelengths (low energies) due to the shift of the plasmon frequency below the visible region. The FOM of thin films increases from ~$10^{-1}$ (x = 0) to $10^1$ (x = 0.30) at $\lambda$ = 550 nm which is a 100 times improvement in the performance of the parent transparent conductor SNO (and a ~1000 times improvement at $\lambda$=700 nm). At higher doping concentration (x = 0.5), the sheet resistance of the thin film becomes quite large, still it manages to outperform SVO, ITO, SMO etc. All these studies validate the usefulness and potential of Ti doped SNO thin films for transparent electrodes.

As a final note, let us emphasize again that the film growth conditions, strain, deposition temperature, laser fluence, target to substrate distance, background partial pressure, thickness, etc. have a huge impact on the optoelectronic properties. These variations can even result in a metal to insulator transition (MIT) in strongly correlated systems, can make the films highly transparent or highly absorbing, depending on the use case. For example, the use of substrates like $DyScO_3$/$KTaO_3$/LSAT show a resistivity in the range of 10 µΩ-cm while obtaining ~50-80% transparency[29,42,51,52], whereas for samples grown on LAO under different conditions SNO films exhibit a resistivity of $10^2$-$10^7$ µΩ-cm and transparency of 40-90%[39,43]. Therefore, to properly compare the optoelectronic properties of these films, we have kept the same conditions for each film and have not performed individual optimization for each doping concentration. However, in our previous work, we demonstrated that the optoelectronic properties of SNO thin films are also highly dependent on $pO_2$[43]. To verify the effect of $pO_2$ on $SN_{1-x}T_xO$ thin films, we further grew films at different background partial pressures to show the variation in the optoelectronic properties. This partial optimization is displayed as stars in Figure 1(c) and reveals that the sheet resistance (as well as optical transparency) can be further tuned to the application of choice. Moreover, the trend of increasing FOM by Ti doping is maintained across the entire doping range. The complete list of samples, with their sheet resistance and FOM is given in Figure S6 of the supplementary information. It shows that we can further enhance the FOM by performing individual optimization (i.e. by varying thickness, change of substrate, growth temperature, or changing) for every concentration.

**Discussion and Conclusion**

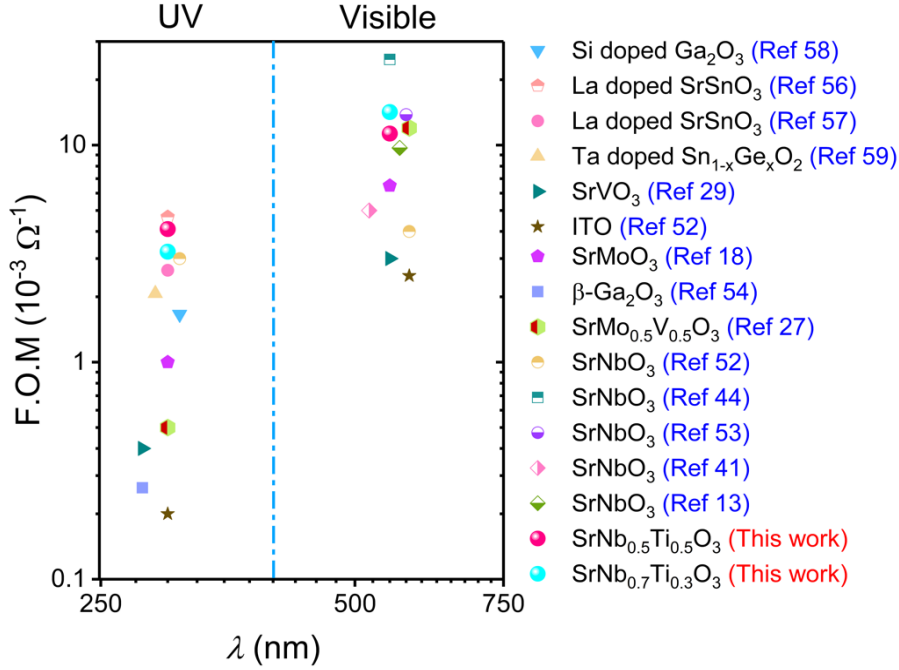

Figure 8: Best achieved FOM obtained for $SN_{1-x}T_xO$ (x = 0.3 and 0.5) thin films compared to other commonly studied transparent materials in the visible and UV spectrum showing that [13,18,27,29,41,44,52–59].

In this work, we explored the unconventional design strategy to enhance the optical transparency of the strongly correlated material $SrNbO_3$ by Ti doping and studied its potential as a new TCO, using a combined theoretical and experimental approach. The design principle employed here is focused on reducing the amount of carrier concentration through Ti doping, which is in contrast to conventional doping that focuses on inducing electronic conductivity in wide-band gap oxides through a Moss-Burstein shift. We observed here that Ti doping reduces the plasma frequency $\omega_p$, shifting the reflection and absorption edge below the visible range. This boosts the transparency of $SN_{1-x}T_xO$ from ~50% to 95% while keeping the sheet resistance below 100Ω, one of the basic requirements of a TCO. Our $SN_{1-x}T_xO$ films also show a remarkable improvement of optical transparency in the ultraviolet (UV) and near infrared (NIR) range of the spectrum, making it highly transparent across the whole UV-Vis-NIR spectrum.

In Figure 8, we compare the FOM of $SN_{1-x}T_xO$ (x = 0.3 and 0.5) thin films with other widely studied transparent conductors. It is seen from the figure that $SN_{1-x}T_xO$ outperforms ITO in both the visible and UV spectrum. This makes $SN_{1-x}T_xO$ a particularly attractive option for solar cells, which requires high transparency across a large electromagnetic spectrum. In the visible range, it also outperforms the recently revealed series of strongly correlated materials (SVO, SMO, $SV_{0.5}M_{0.5}O$ etc.) as well as other reported values of SNO[13,18,27,29,52,52,53]. The FOM can be further optimized by an appropriate choice of the substrate and should then even surpass the record FOM achieved by Jeong et al.[44] for our parent compound SNO. Our calculated FOM in the UV spectrum is 3.1 and 3.9 ($\times 10^{-3}$ $\Omega^{-1}$) for x = 0.3 and 0.5, which is comparable to La doped (Ba, Sr)SnO$_3$ reported recently as a promising TCO in the UV range[55–59]. Figure 8 clearly shows that Ti doped SNO system outperforms other materials as it covers a wider electromagnetic spectrum.

In conclusion, we shed light on hole doping of strongly correlated systems and find that it is an excellent way to enhance the optoelectronic properties. This strategy is straightforward and requires little to no optimization while boosting the FOM. The growth of thin films via other commonly used methods such as sputtering and molecular beam epitaxy (MBE) will further increase the FOM. Thus Ti-doped SNO bears excellent prospects for adaptation in industry. The good agreement of our DFT+DMFT and experimental results emphasizes the power of such a combined approach for identifying new materials for optoelectronics.

**Acknowledgements**

The research funding from Shiv Nadar Institution of Eminence (Deemed to be University) (Grant No. SNS/PHY/2013-20) and DST-Science and Engineering Research Board (SERB)

India (Grant No. SR/FST/PS-I/2017/6C) and the Austrian Science Funds (FWF) through P 36213 is acknowledged. The high-performance computing facility offered by the School of Natural Sciences at Shiv Nadar Institution of Eminence (Deemed to be University) is used for theoretical calculations; DFT+DMFT calculations have been done on the Vienna Scientific Cluster (VSC). L. S. is supported by the National Natural Science Foundation of China (Grants No. 12422407). We acknowledge Prof. S. Dhar from IIT Bombay for performing HR-XRD measurements.

**Methods:**

(a) *Epitaxial growth of thin films:* High purity powders of $SrCO_3$ (Alfa Aesar, 99.9%), $Nb_2O_5$ (Alfa Aesar, 99.9985%) and $TiO_2$ (rutile, Alfa Aesar 99.9%) are mixed in proper molar ratio to get an effective Ti doping of 0, 15, 30, and 50%. These powders are then hand grounded using mortar pestle and calcined at 1473 K till we get a single orthorhombic phase of $Sr_2Nb_{2-x}Ti_xO_7$ (x = 0 to 1). These are then pressed into pellets by applying a pressure of 20 metric tons using hydraulic press in a die set, before calcination in air at 1623K for 96 hours with intermediate grinding every 24 hours. These pellets are subsequently loaded in the PLD chamber (Neocera, USA) to be used as targets for the growth of thin film of appropriate concentration. Next, double side polished single crystal substrates of $LaAlO_3$ (001) (MTI, USA) are loaded into the chamber. Once the base pressure drops to $1\times10^{-7}$ mbar we begin the thin film deposition by setting Oxygen partial pressure ($pO_2$) of $1\times10^{-5}$ mbar by supplying gas through the mass flow controller (MFC). The substrate is then heated to 1023 K in the same background pressure before starting the pulsed laser deposition (Coherent Excimer KrF laser, $\lambda$ = 248nm, Coherent GmbH, Germany). The target to substrate distance is kept constant at 50 mm. The laser fluence chosen for our deposition is 0.8 $Jcm^{-2}$ at a repetition rate of 5Hz. The deposition for each set of doped films takes place under the same conditions until we get thin films of the desired thickness. The sample is finally allowed to cool to room-temperature in the same background pressure as it was fabricated and taken out of the chamber for characterization.

(b) *Film Characterization:* The structural properties of the thin films are measured by high resolution X-ray diffraction at the Industrial Research and Consultancy Centre at IIT Bombay using a Rigaku Smart Lab diffractometer equipped with Cu K$_\alpha$ radiation ($\lambda$ = 1.54 Å). A $\vartheta - 2\theta$ and $2\theta_\chi$ scan is conducted over a range of 20° to 80° with a step increment of 0.001°. The electrical properties of the thin films to determine resistivity, Hall mobility, and carrier concentrations are characterized with the help of physical property measurement system (Quantum Design Inc., USA). The resistance measurements are performed in van der Pauw geometry. The optical transmittance of the thin films is studied by ultraviolet-visible-near infrared spectrophotometer (Shimadzu Solidspec -3700, Japan). The thickness of the thin films and the dielectric functions are determined *ex-situ* using variable angle spectroscopic ellipsometry (VASE) (M-2000-UI, J. A. Woollam, USA). For fitting the data, we have used a combination of Drude and Lorentz oscillator models in CompleteEASE software till the root mean square error between our model and raw data is less than 5.

(c) *DFT simulation:* We perform first-principles calculations with the Vienna ab initio simulation package (VASP) based on projector augmented wave (PAW) pseudopotentials[60,61]. The atomic and electronic structure are investigated by employing the Perdew-Burke-Ernzerhof (PBE) exchange correlation potential[62]. To simulate the effect of Ti doping on SNO, we construct a 2×2×2 supercell of SNO and replace 0, 2, and 4 Nb ions to effectively get 0, 25, and 50% Ti doped SNO using a dense k-mesh of 7×7×7. The lattice constants were set to the experimental values. We choose a cut-off kinetic energy of 600 eV, which is 1.5 times the energy cutoff for O-ions. The bond lengths are relaxed until the energy convergence requirement of $10^{-6}$ eV with Hellman-Feynman forces on each atom at $10^{-3}$ eV/Å is met. To improve total energy accuracy, a gamma-centered Monkhorst-Pack grid with 0.02 Å resolution was adopted. Once the bond length and angles are optimized, we perform band unfolding methods as implemented in vaspkit[63].

(d) *DMFT simulation:* The DFT-level electronic band structures and optical properties of SrNb$_{1-x}$Ti$_x$O$_3$ are recalculated using Wien2k[64–66] with the PBE version of the generalized gradient approximation[62] and the mBJ potential[67]. A dense k-mesh of 13×13×13 (7×7×7) for the 2×2×2 supercell of (doped) SrNbO$_3$ ensures convergence. As before, Ti atoms are introduced into a 2×2×2 supercell of SrNbO$_3$ to simulate doping. All lattice constants and atomic positions are obtained from structural relaxations performed using the VASP code[60,68,69]. The *d*-bands of the doped Ti and $t_{2g}$ orbitals of Nb are wannierized[70,71] using wien2wannier[72,73] and supplemented by local density-density interactions with standard values[74–76] of intra-orbital U = 5.5 eV, inter-orbital V = U-2J = 3.5 eV, and Hund's exchange J = 1.0 eV for Ti 3d-orbitals; as well as 3.0 eV, 2.4 eV, and 0.3 eV for Nb 4d-orbitals. The Nb orbitals are excluded from Wannier projections and subsequent DMFT calculations as they are far above the Fermi level. The generated Hamiltonian is solved at room temperature (300K) using continuous-time quantum Monte Carlo in the hybridization expansion[77] as implemented in w2dynamics[78,79]. Spectra are analytically continued with the maximum entropy method[80,81]. Optical conductivities, and other optical properties such as reflectivity and absorption spectra are calculated using the WOPTIC code[82].

# Supplementary Information for

# Boosting the transparency of metallic SrNbO$_3$ through Ti doping

**Contents:**

**Figure S1:** Band structure and density of states for SrTiO$_3$

**Figure S2:** Comparison of experimentally obtained absorption coefficient with DFT and DMFT results

**Figure S3:** In plane XRD spectrum, $a_{oop}$ and $a_{ip}$, rocking curve, and phi scan of thin films

**Figure S4:** Tauc's plot for calculation of band gap.

**Figure S5:** Ellipsometry data for calculation of plasmon frequency

**Figure S6:** Sheet resistance and FOM for thin films as a function of $p$O$_2$

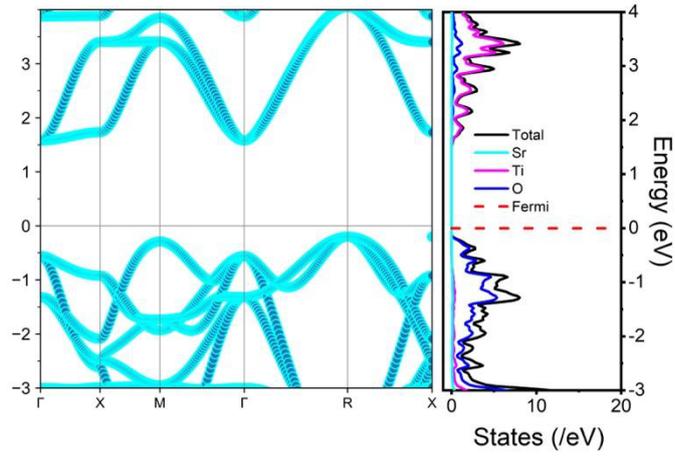

Figure S1: Band structure of pure STO which is an insulator with the Fermi level in the band gap.

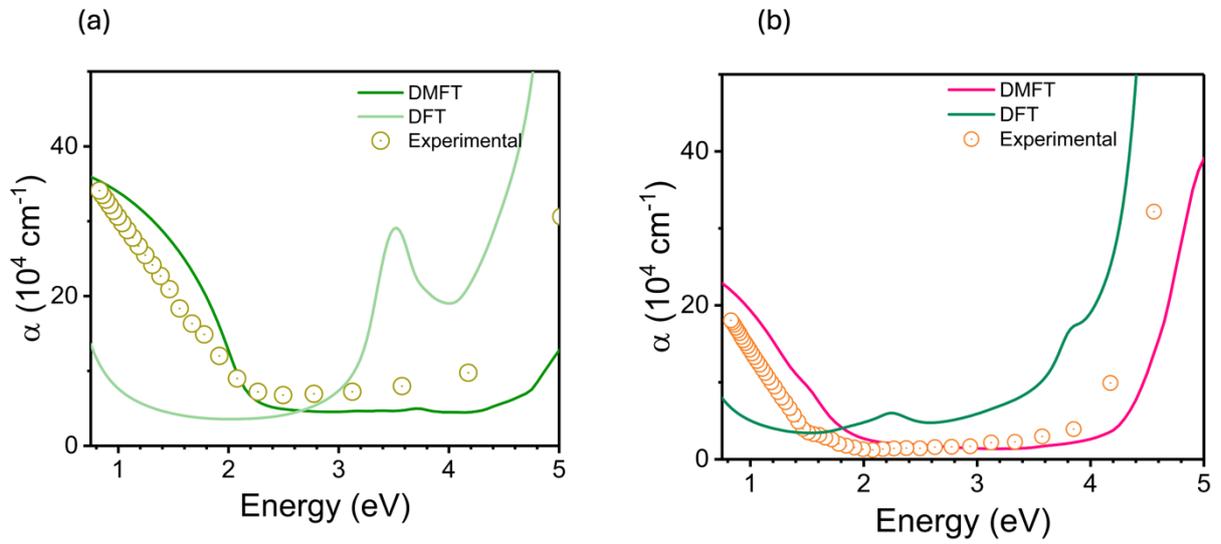

Figure S2: Comparison of DFT and DMFT obtained absorption spectra for (a) SNO and (b) $SN_{0.5}T_{0.5}O$ with the experimental results.

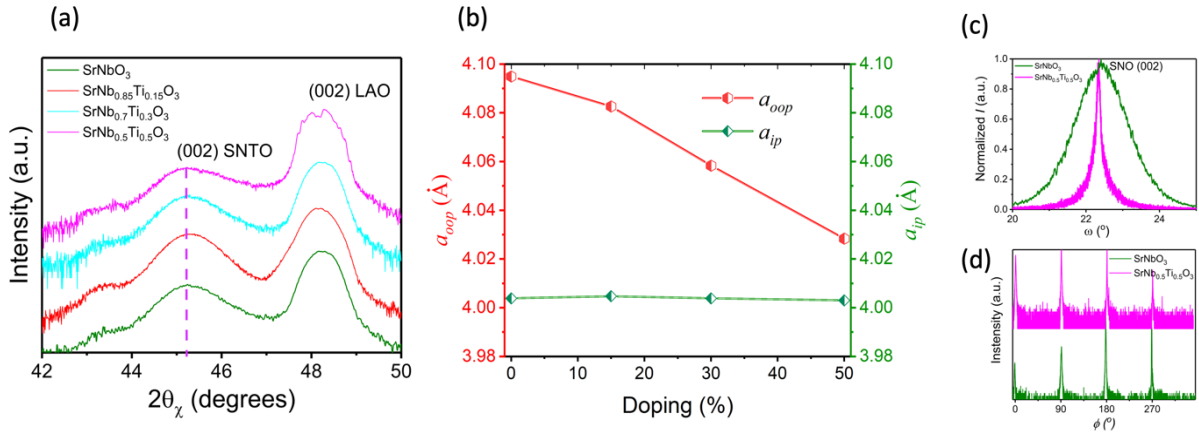

Figure S3: (a) In-plane $2\theta_\chi$ scan (b) calculated in-plane and out-of-plane lattice parameter (c) Phi scan around (103) peak of $SN_{1-x}T_xO$ (x= 0 and 0.5) (d) Variation in the rocking curve around (002) peak for x = 0 and 0.5.

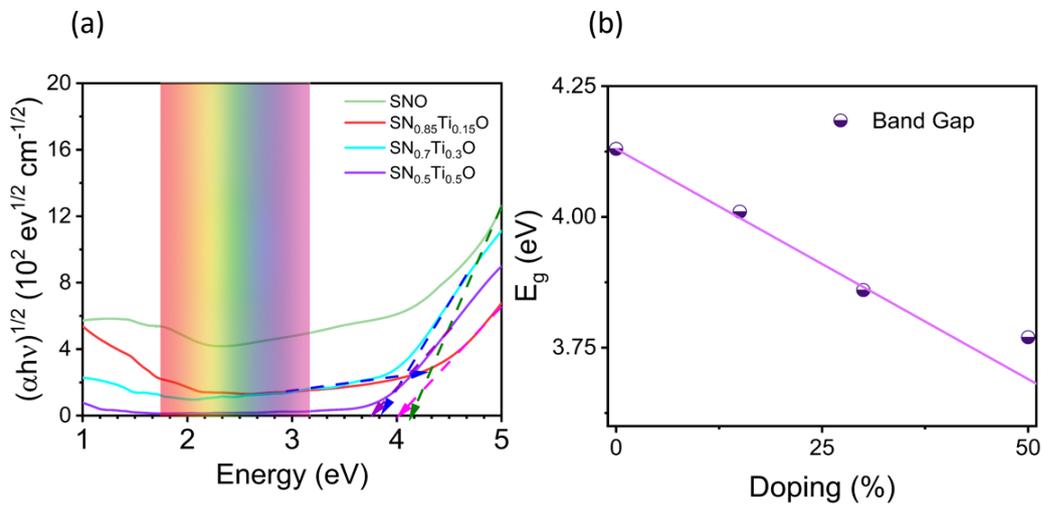

Figure S4: (a) Tauc's plot for calculating the indirect band gap. (b) Calculated band gap of the $SN_{1-x}T_xO$ thin films as function of Ti doping.

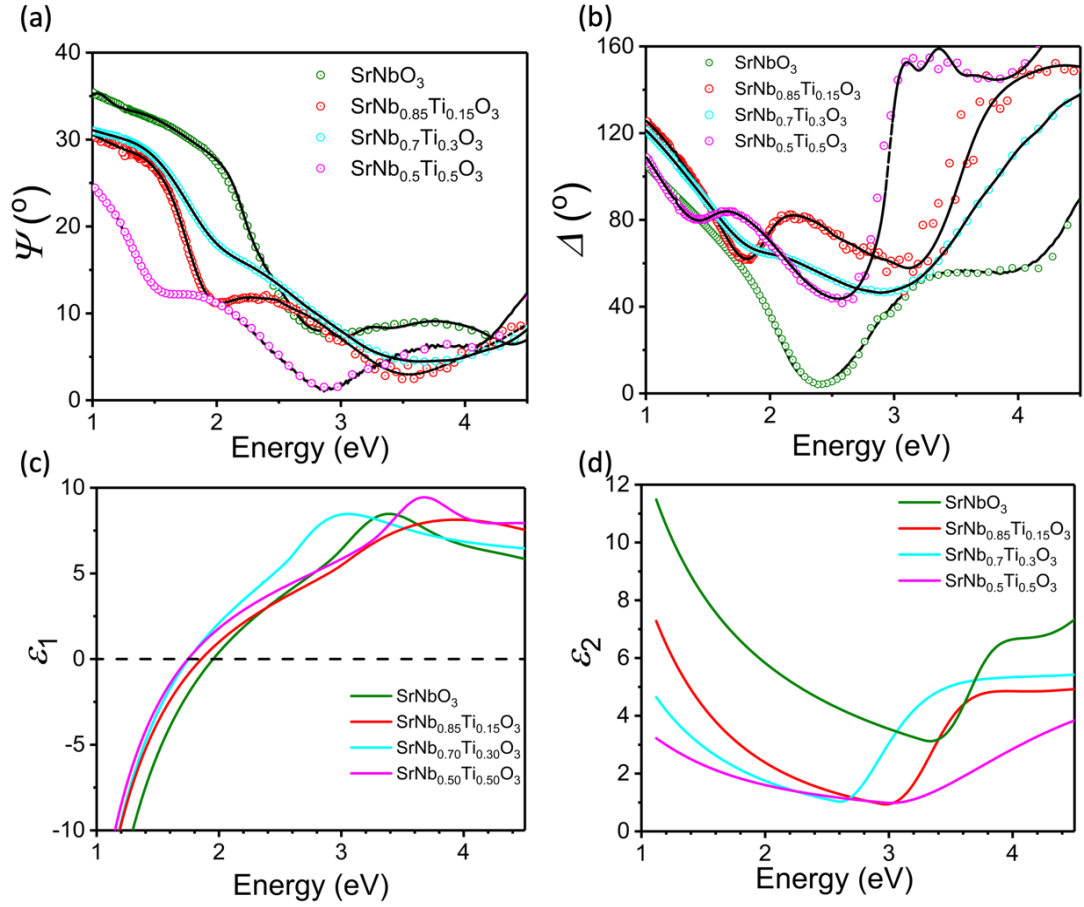

Figure S5: Representative (a) $\Psi$ and (b) $\Delta$ of the ellispometric data at 65° angle of incidence. The open circles represent the raw data, and the black solid line represent the fitting model. (c) Real and (d) imaginary part of the extracted dielectric constants by ellipsometry.

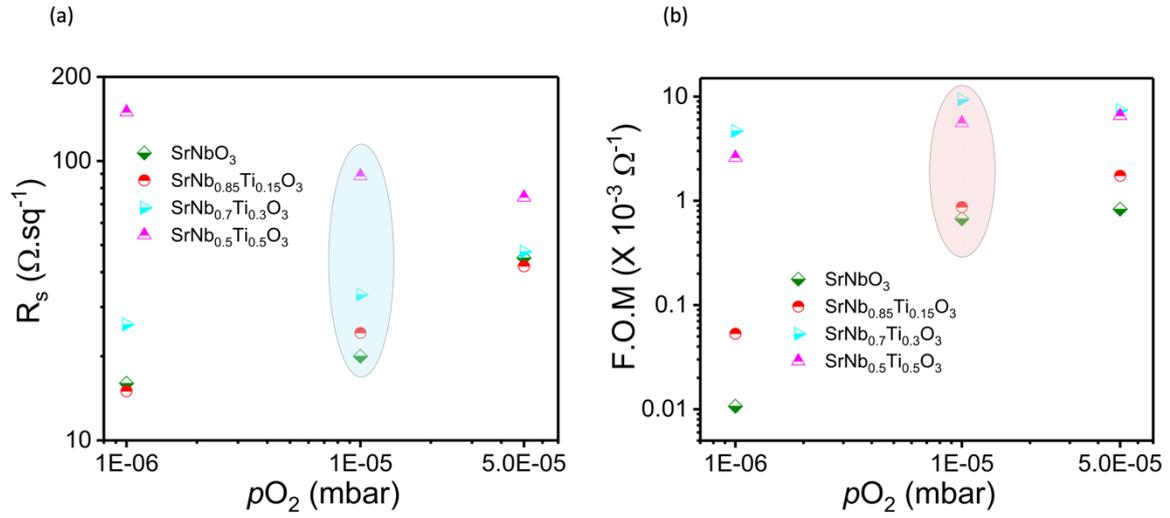

Figure S6: (a) Sheet resistance and (b) figure of merit obtained for thin films grown at different oxygen partial pressure. The shaded portions in each curve represent the growth conditions we have chosen in the main manuscript.